\documentclass{emulateapj}
\slugcomment{Draft Version Jul. 17, 2010: Accepted for Publication in ApJ}
  
\usepackage{times}
\usepackage{bm}

\newcommand{\bolB}{{\bm  B}}

\newcommand{\bolV}{{\bm  V}}

\newcommand{\bolQ}{{\bm  Q}}
\newcommand{\bolF}{{\bm  F}}

\newcommand{\beq}{\begin{equation}}
\newcommand{\eeq}{\end{equation}}
\newcommand{\beqn}{\begin{eqnarray}}
\newcommand{\eeqn}{\end{eqnarray}}
\newcommand{\beqno}{\begin{equation*}}
\newcommand{\eeqno}{\end{equation*}}
\newcommand{\beqnno}{\begin{eqnarray*}}
\newcommand{\eeqnno}{\end{eqnarray*}}

\shorttitle{A Magnetohydrodynamic Model of the M87 Jet. I}
\shortauthors{Nakamura, Garofalo, \& Meier}

\begin{document}

\title{ A Magnetohydrodynamic Model of the M87 Jet\\ I: Superluminal
Knot Ejections from HST-1 as Trails of Quad Relativistic MHD Shocks}

\author{Masanori Nakamura\altaffilmark{1}} \affil{Department of Physics
and Astronomy, The Johns Hopkins University, 3400 N. Charles Street,
Baltimore, MD 21218 \\ Space Telescope Science Institute, 3700 San
Martin Drive, Baltimore, MD 21218; nakamura@stsci.edu}

\and

\author{David Garofalo\altaffilmark{2}, David L. Meier} \affil{Jet
Propulsion Laboratory, California Institute of Technology, Pasadena, CA
91109 ; david.a.garofalo@jpl.nasa.gov, david.l.meier@jpl.nasa.gov}

\altaffiltext{1}{Allan C. Davis Fellow}
\altaffiltext{2}{NASA Postdoctoral Fellow}

\begin{abstract}

This is the first in a series of papers that introduces a new paradigm
for understanding the jet in M87: a collimated relativistic flow in
which strong magnetic fields play a dominant dynamical role. Here we
focus on the flow downstream of {\em HST}-1 --- an essentially
stationary flaring feature that ejects trails of superluminal
components. We propose that these components are quad relativistic
magnetohydrodynamic shock fronts (forward/reverse fast and slow modes)
in a narrow jet with a helically twisted magnetic structure. And we
demonstrate the properties of such shocks with simple one-dimensional
numerical simulations. Quasi-periodic ejections of similar component
trails may be responsible for the M87 jet substructures observed further
downstream on $10^{2-3}$ pc scales. This new paradigm requires the
assimilation of some new concepts into the astrophysical jet community,
particularly the behavior of slow/fast-mode waves/shocks and of
current-driven helical kink instabilities. However, the prospects of
these ideas applying to a large number of other jet systems may make
this worth the effort.
\end{abstract}

\keywords{galaxies:individual: M87 --- galaxies: active --- galaxies:
jets --- methods: numerical --- MHD}

\section{Introduction}
\label{sec:int}

M87 is a nearby giant elliptical galaxy (Virgo A, 3C 274, NGC 4486)
located at the center of the X-ray-luminous Virgo cluster \citep[]{F80},
host of the first extragalactic jet discovered \citep[]{C18}. The
one-sided jet emerging from the nucleus of M87, where a $3.2 \, (\pm0.9)
\times 10^{9}$ solar mass black hole\footnote{Recently, a new black hole
mass of $6.4 \, (\pm0.5) \times 10^{9} M_{\odot}$ has been proposed by
\citet[]{GT09}; however we use $3 \times 10^{9} M_{\odot}$ throughout
the paper.} resides \citep[]{M97}, has been well-studied on a wide range
of wavelengths from radio to X-rays \citep[]{O89, B95, P99, B99, J99,
M02, WY02, PW05, H06, L07, KOV07}. Because of its proximity
\citep[$D=16$ Mpc;][]{T91}, which gives a linear scale of 78 pc
arcsec$^{-1}$, the M87 jet is one of the best candidates to investigate
relativistic outflows in extragalactic systems.

One of the most remarkable features of the M87 jet is the innermost
bright knot G, lying about 1$\arcsec$ from the core in the {\em Very
Large Array} (VLA) observations \citep[]{O89}. That region has been
resolved by {\em Hubble Space Telescope} (HST) into a structured complex
known as {\em HST}-1 \citep[]{B99}. It is located around 0.8-1$\arcsec$
(projected) from the core (or $\sim 260 - 320$ pc de-projected for a
viewing angle of $\sim 14^{\circ}$; Wang \& Zhou 2009). It appears
almost stationary, but component velocities downstream of the complex
are highly relativistic, with a range $4c - 6c$ \citep[]{B99}. Indeed,
the component of {\em HST}-1 that is furthest upstream (i.e., {\em
HST}-1d) is stationary to within the errors ($<0.25 \ c$), and has been
identified as the point of origin of the superluminal ejections
\citep[]{C07}. As of this writing, no other observations have detected
any superluminal components upstream of the {\em HST}-1 complex.
Therefore, the observations obtained up to now, paint a surprising
picture: no evidence for highly relativistic velocities between the core
and {\em HST}-1, a stationary knot at {\em HST}-1d, and then, suddenly,
superluminal motion of components immediately downstream of {\em HST}-1.

The structure of the jet downstream of {\em HST}-1 (1 - 18$\arcsec$ or
0.1 - 1.5 kpc in projected distance) can be characterized by trailing
clumps or knots of bright gas ({\em HST}-1 to C) with an apparent
deceleration to subluminal speeds (around $6c$ to $0.5c$) \citep[]{B95,
B99} and filamentary structures (``wiggles/kinks'') \citep[]{O89,
S96}. The high overpressure in the synchrotron gas and the highly
polarized helical filaments \citep[]{O89, P99} indicate the existence of
underlying ordered magnetic fields with a 3-dimensional helix seen in
projection; magnetic fields, therefore, appear to play a role in
determining the M87 jet structure even on large scales. The magnetic
field vectors in the knots {\em HST}-1, D, A, and C are perpendicular to
the jet direction, indicating the presence of longitudinal compression
by a shock front and/or a tightly wound magnetic helix \citep[]{O89,
P03}. Detailed broadband (from radio through optical to X-ray) spectral
shape of the knots ({\em HST}-1 to C) in the M87 jet favors the scenario
in which synchrotron emission dominates the radiation and {\it in situ}
particle acceleration (by the first-order Fermi process) almost
certainly occurs in the large scale M87 jet (both within knots and
outside them) \citep[]{PW05}.

The purpose of the paper is to investigate the dynamics downstream of
{\em HST}-1 using relativistic magnetohydrodynamics (MHD). We introduce
here the concept that the superluminal components in M87, and perhaps
extragalactic jets in general, are the result of MHD shocks produced in
helically twisted, magnetized relativistic outflows. Our exclusive focus
here on the flow just downstream of {\em HST}-1, a unique dynamical
behavior of a pair of sub/super-luminal knots, constitutes the first in
a series of papers on the MHD paradigm for the entire M87 jet. The paper
is organized as follows. We outline our model in \S \ref{sec:Model}. In
\S \ref{sec:Method} our numerical method is introduced, and in \S
\ref{sec:Results} we show our numerical results. Discussions and
conclusions are given in \S \ref{sec:Discussion}.

\section{Model Description}
\label{sec:Model}

We propose an MHD model of the M87 jet that extends beyond, or
downstream, of {\em HST}-1 that resolved by the VLA and HST observations
($1 - 18 \arcsec$ or $0.1 - 1.5$ kpc in projection).  The structure of
the jet downstream of {\em HST}-1 can be characterized by a conical
shape $z \propto r$ (where $r$ is a jet radius and $z$ is a distance
from the nucleus) with an opening angle of $\theta_{\rm open} \sim
6^{\circ}$ \citep[]{VN79, O80, R82, O89}. The high degrees of radio and
optical polarization in both the knot ($40\% - 60\%$) and interknot
($20\% - 40\%$) regions \citep[]{P99} indicate the presence of coherent
magnetic fields (highly ordered) on large scales associated with the
underlying jet.

Under the assumption of the minimum energy condition, the knots
themselves appear to be significantly overpressured \citep[]{O89} with
respect to the ambient thermal gas \citep[]{YWM02}, {\rm but} the
interknot regions do not \citep[]{S96}. In order to maintain a conical
streamline of the adiabatic jet within a uniform ambient gas (an
iso-thermal King profile with a core radius $R_{\rm c} \simeq 18\arcsec
$), the fields may have to be much stronger and more highly ordered
({\em i.e.}, a force-free configuration with a 3-dimensional helix) than
a weak and tangled field at the equipartition level ($\sim$ a few of
100\,$\mu$G).  Magnetic fields, therefore, appear to play a crucial role
in determining the structure of the M87 jet even on larger scales over
100 pc.

In the present paper, we do not examine the entire three dimensional
structure of the M87 jet, described in \S \ref{sec:int}; this will be
treated in forthcoming papers. Instead, we focus on the unique
properties of the ejected superluminal components \citep[]{C07} by using
a simple one dimensional approach. \citet[]{C07} reported that some time
between 2005 December and 2006 February, one of the newly ejected
components, called {\em HST}-1c, {\em split into two bright features}: a
faster moving component (c1: $4.3c \pm 0.7c$) and a slower moving (c2:
$0.47c \pm 0.39c$). If the bright knot is a shock, then such a pair of
separated components naturally can be identified as forward/reverse
modes. We here model this interesting feature by relativistic MHD
simulations. In addition, we suggest that a similar expansion might be
occurring in earlier {\em HST} observations ({\em HST}-1$\epsilon$:
$6.00c \pm 0.48c$ / {\em HST}-1 East: $0.84c \pm 0.11c$) by
\citet[]{B99}.

\begin{figure} 
\begin{center}
\includegraphics[scale=0.27, angle=0]{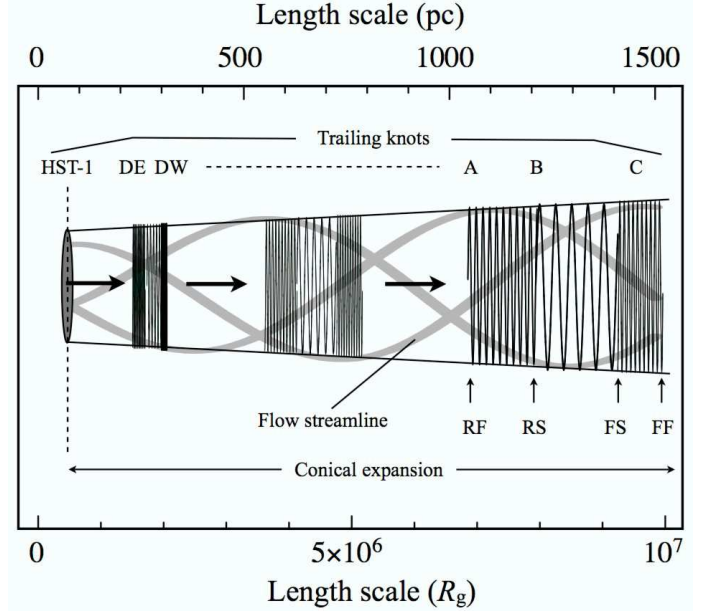} 
\caption{ \label{fig:f1}
Schematic view of our model of the 87 jet in the VLA scale. The
structure of the jet downstream of {\em HST}-1 can be characterized by
trailing MHD quad shocks that may be intermittently generated at the
{\em HST}-1 complex. The system of MHD quad shocks consists of the
forward fast (FF) and slow (FS) followed by the reverse slow (RS) and
fast (RF) shocks.}
\end{center}
\end{figure}

Furthermore, the Knot D complex located just downstream of the {\em
HST}-1 complex, also displays distinct features (DE -- DM -- DW) with
different propagation speeds and a sideways oscillation of the
brightness ridge-line \citep[]{B95}. It has been pointed out that four
quasi-periodic knot complexes downstream of {\em HST}-1 (called D, E, F,
and I) are at roughly uniform spacing \citep[]{BM93}. Thus, recurrent
ejections of superluminal components at the {\em HST}-1d site may happen
with a certain quasi-periodic time scale. We suggest that these bright
components can be considered as similar paired structures to {\em
HST}-1c2/c1 and {\em HST}-1 East/$\epsilon$: DE -- DM -- DW, E -- EF, F
-- I, and A -- B -- C further downstream [refer to \citet[]{O89} and
\citet[]{B99} for definitions of the know labels].

The stationary knot in {\em HST}-1d may be identified as the
recollimation shock \citep[]{S06, BL09}. Once the recollimation shock is
formed at a finite distance, the reflection of supersonic flows may emit
a shock component downstream of the recollimation point. A Steady model
of relativistic MHD multi-layer outflows (relativistic jet and
non-relativistic wind) by \citet[]{G09} nicely reproduces such observed
properties of M87 as an asymptotic collimation \citep[]{J99} with a
feature brightened at around 100 pc (projected distance) by an
over-collimating MHD flow.

Of particular interest is the brightest emission structure A -- B -- C.
As reported by \citet[]{O89} and \citet[]{P99}, knots A and C have
certain similarities: i) bright transverse linear features (normal to
the jet axis) indicative of a shock front \citep[]{B83}; ii) dominance
of transverse magnetic field suggesting ordered helical magnetic
components. Visible side-to-side oscillation is also observed between
these knots, and magnetic vectors appear to follow the fluctuating jet
axis in this part (including knot B).

We point out a sudden enhancement of emission at knot A, at the {\em
upstream} edge of that knot, indicating a reverse shock and a rapid drop
in emission at the {\em downstream} edge of knot C, suggesting that it
is the corresponding forward shock \citep[see, Figure 8a, 10a
of][]{O89}. Particle acceleration is associated with both the forward
and reverse modes. In addition, the substructure between the upstream
edge of knot A and the downstream edge of knot C indicates there may be
another pair of (forward/reverse) shocks or rarefaction waves. The
downstream edge of knot A or knot B may be a reverse mode feature, while the
upstream of knot C may be a forward mode feature.

Sudden changes in magnetic vector orientation strongly imply the
existence of MHD fast/slow mode waves. The transverse component of the
magnetic field $\bolB$ ($B_{\perp}$) {\em increases} across a fast mode
front (the normal component $B_{\parallel}$ remains unchanged through an
MHD oblique shock and thus the magnetic pressure $p_{\rm m}$ increases),
while $B_{\perp}$ {\em decreases} across a slow mode front ($p_{\rm m}$
decreases). Thus, we interpret the brightest emission structure A -- B
-- C to be a trail of {\em quad} relativistic MHD shocks (forward
fast/slow and reverse slow/fast) (perhaps, two slow modes can be
rarefaction waves) generated in a helically twisted super-fast magnetic
flow. Such a flow can be generally expected at further downstream from
the AGN in the MHD jet theory and emissions of MHD shocks from the
stationary knot feature has been inspected by MHD simulations of
strongly magnetized jets \citep[]{L89}. The presence of four shocks,
{\em instead of two}, is due to the presence of $B_{\phi}$, which serves
a role equivalent to $B_{\perp}$ in the simpler planar oblique shock
case.

The transverse field component is dominant in some knots (HST-1, D, A,
and C), indicating strong longitudinal compressions by a passing
shock. In the past, a transverse orientation of the magnetic field was
thought to be due to longitudinal compression of roughly random
(tangled) magnetic fields ({\em weakly polarized}) by a
hydrodynamic-like shock \citep[]{L80} ({\em i.e.}, this is essentially a
{\em perpendicular} shock). However, as is shown in \citet[]{P99}, the
high degree of polarization is confirmed in interknot regions of the M87
jet on scales of $10^{2-3}$ pc. So the classical picture of a weak,
random jet magnetic field may be in conflict with these observations. It
is natural to conclude, therefore, that underlying magnetic fields in
both the interknot and knot regions are systematically ordered with
helical (longitudinal + azimuthal) components (three-dimensional helical
structure). Longitudinal compression of a helical magnetic field also
will produce an enhanced azimuthal (i.e., projected transverse)
component.

High energy particle acceleration by the first-order Fermi process is
generally believed to occur in extragalactic jets \citep[]{BO78}.  A
relativistic particle energy distribution [$n(E)\propto E^{-\delta}$]
steeper than $\delta=2$ would be needed to produce the radio - optical -
X-ray synchrotron spectrum in the M87 jet.  Synchrotron model fits from
radio through optical to X-ray data produce $\delta=2.2$ at all energies
and all locations along the jet \citep[]{PW05} and about $\delta=2.36$
on average \citep[]{LS07}.  These agree very well with the conditions
needed for a diffusive shock acceleration (DSA) [$\delta=2-2.5$;
\citet[]{KD01}].

The geometry of a perpendicular MHD shock (this is a so-called
``magneto-acoustic shock'', a particular case of the fast-mode MHD
oblique shock with only magnetic component along the shock surface
$B_{\perp}$) is not suitable for a DSA; negligible particle acceleration
has been confirmed by PIC simulations \citep[]{L88, G92}. Furthermore,
in the standard picture of Fermi acceleration, the DSA does not work for
slow-mode MHD shocks because the magnetic field strength decreases
across them, and therefore they cannot act as magnetic mirrors for the
upstream particles. Thus, {\em only fast-mode MHD oblique shocks are
probably responsible for the observed non-thermal emissions}.

In summary, we suggest that each pair of bright knots on $10^{2-3}$ pc
scales in the M87 jet is a quasi-periodic event produced in the
stationary {\em {\em HST}-1}d knot as a consequence of recollimation
processes of a converging super-fast magnetosonic jet. They are trailing
quad MHD shock wave fronts (forward fast/slow modes and reverse
slow/fast modes), which propagate as super/sub-luminal components in a
{\em highly magnetized and twisted}, relativistic outflow powered by a
non-linear torsional Alfv\'en wave train \citep[]{MEI01}. A schematic
view of our M87 model is shown in Figure \ref{fig:f1}.

The physics of MHD shocks in helically twisted magnetized flows are
still not very well known in the astrophysical jet community. Despite
the fact that jet dynamics is inherently three-dimensional, much of the
physics of these jets can be obtained from simple 1-D simulations of
flow along a cylindrical shell. Therefore, in this paper we investigate
observed properties of {\em HST-1}c \citep[]{C07} using a simple
one-dimensional approach that suffices in illustrating the basic
principles. We demonstrate that observed proper motions of forward
(superluminal) and reverse (subluminal) knots can be reproduced
precisely in our relativistic MHD simulation model.

\section{Numerical Method}
\label{sec:Method} 

Based on observations \citep[see, {\em e.g.},][]{O89, P99}, we assume
that the magnetic field plays a fundamental role in determining the flow
properties of the M87 jet all the way up to scales beyond $100$ pc in
projection. We impose axisymmetry and choose a cylindrical coordinate
system $(r,\ \phi,\ z)$ whose axis coincides with the symmetry-axis. We
model the dynamical behavior of observed {\em HST}-1c knot which split
into two distinct features as sub/super-luminal knots \citep[]{C07}.  We
here consider initial phase of the knot ejection and separation
(propagation of individual knots with constant speeds) and thus dynamics
of the flow can be described in so-called 1.5-dimensional approximation
along a cylindrical shell of radius $r_0$ (which is assumed to be
rigid), which allows the quantities to vary in the $z$-direction, and
which also allows for the influence of azimuthal effects.

Our system obeys ideal, special relativistic MHD \citep[]{L67} that
consists of the baryon mass and energy-momentum conservation laws in the
absence of a gravitational field, Maxwell equations in CGS-Gaussian
units, and Ohm's law: 
\beqn 
&&\frac{\partial \bolQ}{\partial
t}+\frac{\partial \bolF}{\partial z}=0, \\ &&\bolQ \equiv \left[
\begin{array}{c}
\gamma \rho \\
\gamma^2 h V_\phi / c^2 - E_{r} B_{z} / (4 \pi c) \\
\gamma^2 h V_z / c^2 + E_{r} B_{\phi}/ (4 \pi c) \\
B_{\phi} \\ 
\gamma^2 h - p + (E_{r}^2 + B_\phi^2 + B_{z}^2) / (8 \pi)
\end{array}
\right], \\
&&\bolF (\bolQ) \equiv \left[
\begin{array}{c}
\gamma \rho V_{z} \\ 
\gamma^2 h V_{\phi} V_{z} / c^2 - B_{\phi} B_{z} / (4 \pi) \\
\gamma^2 h V_{z}^2 / c^2 + p + (E_{r}^2+B_\phi^2-B_{z}^2) / (8 \pi) \\ 
c E_{r} \\ 
\gamma^2 h V_{z}+c E_{r} B_{\phi} / (4\pi)
\end{array}
\right].
\eeqn
The above equations denote the mass, the azimuthal ($\phi$) momentum,
the axial ($z$) momentum, the azimuthal induction, and the energy
equations respectively.

Here $\rho$ and $p$ are the rest-mass density and gas pressure in the
fluid frame. $\bolV=(V_{\phi},\ V_{z})$ is the fluid velocity and
$\gamma$ is the associated Lorentz factor $\gamma \equiv
1/(1-V^2/c^2)^{1/2}$. $c$ is the speed of light, and $\Gamma = 5/3$ is the
specific heat ratio. $\bolB=(B_{\phi},\ B_{z})$ denotes magnetic field
as measured in the laboratory (galaxy) frame. (Note that $B_{z}$ is
constant by $\nabla \cdot \bolB=0$ and never changes in time.) $h = \rho
c^{2} + \Gamma p /(\Gamma-1)$ is the relativistic enthalpy, and $E_{r}
[=(V_{\phi}B_{z}-V_{z}B_{\phi})/c]$ is the radial component of the
electric field as measured in the laboratory frame.

We normalize physical quantities by unit length scales $L_{0}$, density
$\rho_{0}$, velocity $V_{0}$ in the system, and other quantities derived
from their combinations, {\it e.g.}, time as $L_0/V_0$, etc. This
normalization does not change the form of the basic equations. A factor
of $4 \pi$ has been absorbed into the scaling for both $E_{r}$ and
$\bolB$. The resulting set of time-dependent, fully conservative
equations for Special Relativistic MHD (SRMHD) is solved by a finite
volume method (FVM). We use a newly designed, hybrid flux (HF) scheme
\citep[]{N10} that consists of a unique hybridization of a Godunov-type
and centered difference-based fluxes to achieve low-level numerical
diffusion. (Typical 1-D Riemann problems exhibit $\delta^{\rm HF}/
\delta^{\rm HLL} \approx 0.6$ and $\delta^{\rm HF}/\delta^{\rm Roe}
\approx 0.7$, where $\delta$ is the relative error determined from an
$L_1$ norm). MUSCL-TVD reconstruction \citep[]{VL79} and MUSCL-Hancock
predictor-corrector time marching \citep[]{VL84} schemes, with a van
Albada limiter \citep[]{VA82}, are implemented to maintain second-order
accuracy in both space and time.

\section{Numerical Results}
\label{sec:Results}

We adopt $r_{0}=1$ pc as the unit length $L_{0}$ (and as the radius of
{\em HST}-1d) and $c=3 \times 10^{10}$ cm s$^{-1}$ as unit velocity
$V_{0}$. An electron density $n_e=0.17$ cm$^{-3}$ and a temperature
$kT=0.8$ keV at the Bondi radius $\sim 120$ pc \citep[]{A06} are taken
as central ISM properties. These correspond to a density $\rho_{\rm ISM}
(=\mu m_p n_e) =1.7 \times 10^{-25}$ g cm$^{-3}$ ($m_p$ is the proton
mass and $\mu=0.61$ is the molecular weight for full ionization) and
temperature $T_{\rm ISM}=9.3 \times 10^6$ K. We also assume a unit
density $\rho_0=1.7 \times 10^{-27}$ g cm$^{-3}$, which is two orders
smaller than $\rho_{\rm ISM}$ (light jets are generally believed to be
present in radio galaxies and quasars) \citep[see, {\it
e.g.,}][]{K03}. This gives $\rho_0 c^2=1.5 \times 10^{-6}$ dyn cm$^{-2}$
as the unit pressure. The unit time $t_0$ becomes $r_0/c=10^{8}$ s
$=3.2$ yr, and the unit magnetic field $B_{0}$ is $(4 \pi \rho_0
c^2)^{1/2}=4.3$ mG.

The computational domain $z \in [-0.04,\,2.0]$ (pc in a dimensional
scale) is resolved with 5100 grid points. A Riemann problem, which
possesses two uniform initial states (l: left and r: right) separated at
$z=0.0$, is considered to investigate the evolution of the system.  Our
fiducial model consists of a super-fast magnetosonic, relativistic jet
inflow ($\gamma \simeq 6.22$) into a nearly force-free, weakly twisted
($B_{\phi}/B_{z} \simeq 0.33$) medium flowing with a sub-relativistic
speed ($\gamma \simeq 1.07$). It produces a set of quad relativistic MHD
shocks. The following initial conditions are prescribed:
\begin{eqnarray}
&&(\rho,\,V_{\phi},\,V_{z},\,B_{\phi},\,B_{z},\,p)^{\rm l}=
 (1.0,\,0,\,0.987,\,0.8,\,2.4,\,0.256) \nonumber
\end{eqnarray}
on the left-hand side $(-0.04 \leq z \leq 0.0)$ and
\begin{eqnarray}
&&(\rho,\,V_{\phi},\,V_{z},\,B_{\phi},\,B_{z},\,p)^{\rm r}=
 (1.0,\,0,\,0.36,\,0.8,\,2.4,\,0.256) \nonumber
\end{eqnarray}
on the right-hand side. Time integration, using a CFL number of 0.8, is
followed until $t=2.0$ ($\sim 6.4$ yr) to examine the earliest phases of
relativistic MHD shock propagations.

The above choice of initial values has been inferred carefully from various
observational constraints. By assuming a viewing angle $\theta_{\rm v}
\sim 14^{\circ}$ at {\em HST}-1 \citep[]{WZ09}, a maximum pattern speed
of a faster moving component {\em HST-1}c1 (including an error) can be
estimated from its apparent speed $\beta_{\rm app} = 4.3c \pm 0.7c$
\citep[]{C07} as
\beqn
\beta_{\rm pattern}&=&\frac{\beta_{\rm app}}{\beta_{\rm
app} \cos \theta_{\rm v}+\sin \theta_{\rm v}} \simeq 0.982,
\eeqn
where $\beta=V/c$. It would seem far more natural to have the pattern
speed tied to the jet fluid speed, with $\beta_{\rm pattern} \lesssim
\beta_{\rm fluid}$ \citep[]{B95}. We note $\beta_{\rm pattern}=0.989$
corresponds to $\theta_{\rm v}=19^{\circ}$, an upper limit of the
viewing angle for possible solutions to have the superluminal motion
$\beta_{\rm app}=6.0$ as seen in HST observations \citep[]{B99}.  On the
other hand, an ambient motion in the vicinity of {\em HST}-1 complex can
be constrained by the stationary feature of {\em HST}-1d with $\beta_{\rm
app} < 0.25$ \citep[]{C07} as $\beta_{\rm pattern} < 0.516$. The
magnetic field vectors in projection around {\em HST}-1 are
perpendicular to the jet direction and the timescale of optical and
X-ray variability of {\em HST}-1 requires $|\bolB| \sim 10$ mG
\citep[]{P03}. Taking into account a viewing angle, the constraint for
the projected $\bolB$-vectors to be perpendicular to the jet can be
expressed \citep[]{A08} as 
\beqn 
\frac{B_{\phi}}{B_{z}} > \sin
\theta_{\rm v} \simeq 0.24. 
\eeqn 
Thus, our simulation model is reasonably guided by observations. In the
following, we examine our results.

\begin{figure} 
\begin{center}
\includegraphics[scale=0.19, angle=0]{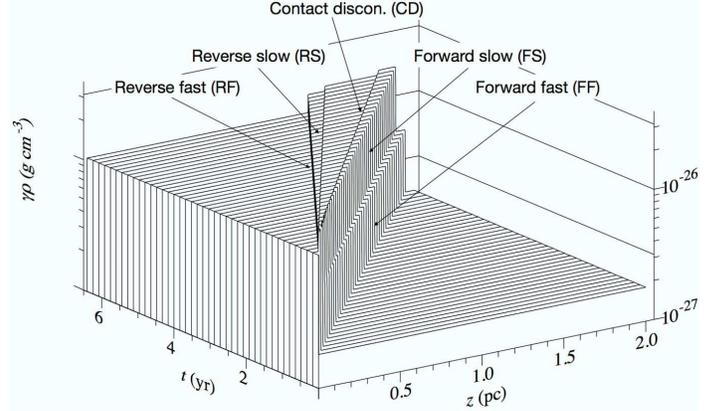} 
\caption{\label{fig:f2}
Space-time ($z$, $t$) diagram of logarithm of the proper density $\gamma \rho$ (vertical 
axis), as measured in the laboratory (galaxy) frame. Quad MHD shocks; FF, FS, RS, and 
RF, and a contact discontinuity (CD) are labeled.}
\end{center}
\end{figure}

Figure \ref{fig:f2} shows the propagation of MHD wave fronts in the
proper density $\gamma \rho$. Quad MHD shocks and a contact
discontinuity (CD or entropy wave), all with constant speeds, are
clearly visible. Relative to a reference frame that co-moves, {\em and
co-rotates}, with the jet plasma near the CD, these waves propagate in
both the forward (F) and reverse (R) directions. Here we adopt the
convention of counting shocks beginning with the one farthest from the
origin of the disturbances ({\em HST}-1). Two of the four shocks, the
first and the fourth, are forward fast-mode (FF) and reverse fast-mode
(RF) shocks, respectively. The other two, the second and the third, are
forward slow-mode (FS) and reverse slow-mode (RS) shocks.

\begin{figure} 
\begin{center}
\includegraphics[scale=0.47, angle=0]{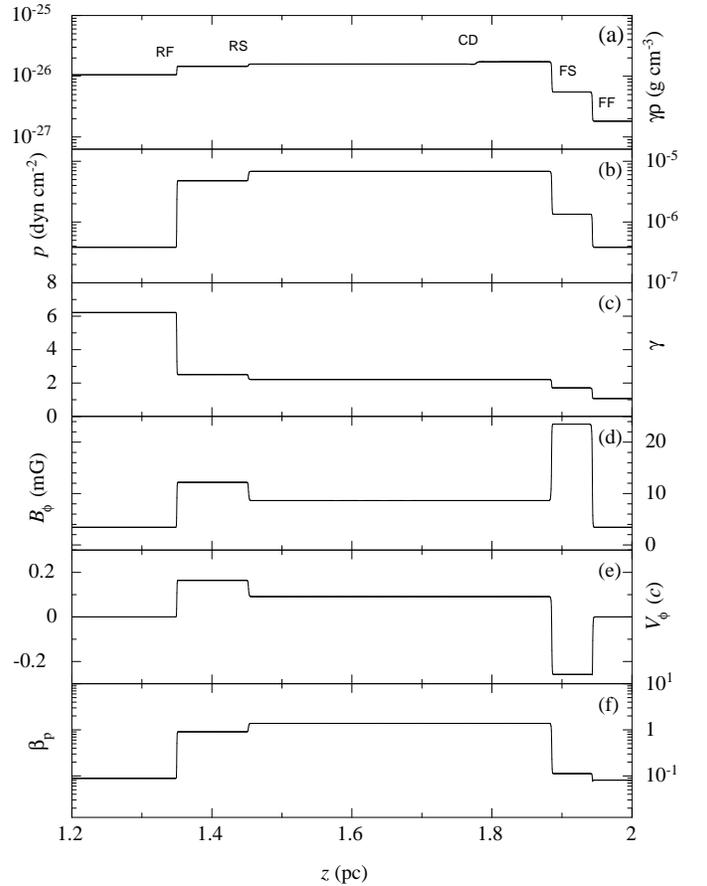} 
\caption{ \label{fig:f3} (a)-(f) $\log (\gamma \rho)$, $\log (p)$,
$\gamma$, $V_{\phi}$, $B_{\phi}$, and $\beta_{\rm p}$ (plasma-$\beta$),
respectively, shown at $t = 2.0$. Only the region $1.2 \leq z \leq 2.0$
is displayed. Note that panels (a)-(e) are measured in the rest frame of
the galaxy, but the panel (f) is measured in the rest frame of fluid
element. Each discontinuity is labeled in (a) (see also Fig. 
\ref{fig:f1}).}
\end{center}
\end{figure}

Snapshots of various quantities at $t = 2.0$ are illustrated in Figure
\ref{fig:f3}. As seen in (a) and (b), the gas is compressed across the
first (FF) and second (FS) shocks. In crossing the third shock (RS) the
gas is expanded by a smaller ratio than the FS, while it is expanded
much more strongly in crossing the last (RF) shock rather than the FF as
clearly shown in (b) [it may not be visible in (a), but this is due to
the frame of the measurement; in the rest frame of fluid elements, the
distribution of $\rho$, instead of $\gamma \rho$, has a similar tendency
with $p$]. As a result, the gas accumulates in the region between the
second and third shocks. As one moves from large to small $z$, $\gamma$
increases with gradual steps throughout the first, second, and third
shocks, but largely increases at the last shock towards the injection
level $\gamma \simeq 6.22$ shown in (c). Similarly, from (d), $B_{\phi}$
increases across the first shock, decreases across the second one,
increases again across the third shock, and finally decreases across the
fourth one. From (e), $V_{\phi}$ changes as well, in a way consistent
with the increased twist between the first and the second shocks, and
reversed between the third and fourth shocks.\footnote{Note that the
region FF - FS and the region RS - RF are counter-rotating when viewed
from a frame that rotates with the plasma near the CD.}

We define the plasma-$\beta$ (a ratio of the gas pressure to the
magnetic pressure) in the rest frame of the fluid element:
\beqn
\label{eq:beta}
\beta_{\rm p} \equiv \frac{2 p}{B_z^{2}+B_{\phi}^2/\gamma^2}. 
\eeqn
This enables us to compare the importance of the magnetic forces with
the plasma forces in the proper rest frame of the fluid. As shown in
(f), the unshocked region further downstream of the FF is highly
magnetically dominated $\beta_{\rm p} \simeq 0.08$, which is prescribed
as the initial condition. $\beta_{\rm p}$ increases slightly in crossing
the FF (both the gas and magnetic pressures are enhanced
similarly). However the accumulated gas region (RS--CD--FS) is heated
strongly (the gas pressure increases, while the magnetic pressure
decreases) and thus attains near equipartition ($\beta_{\rm p} \sim
1.4$). At the forward part of the shock quad (FS--FF) $\beta_{\rm p}$ is
still very small ($\sim 0.1$), but at the reverse part (RF--RS) it has
increased by nearly one order ($\sim 0.9$). In the region behind the
shock quad, after the four MHD shocks have passed, $\beta_{\rm p}$
decreases to a low level that is prescribed in the initial condition.
Given the fact that the relevant physical quantity is the fluid frame
value of the plasma-$\beta$ ($\beta_{\rm p}$: eq. \ref{eq:beta}), and
that $B_{\phi}/B_{z} < 1$, $\beta_{\rm p}$ is not appreciably different
from that in the initial values adopted in the galaxy frame. Note that
the relativistic length-contraction effect produces a weak compression
of the structures between FS--FF rather than between RF--RS when viewed
in the laboratory frame (which is also the computational grid on which
the simulation is done).

Strengths and propagation speeds of the four shocks remain constant with
distance as they propagate in our coordinate system: axial propagation
($z$-direction) in a uniform medium (constant sound and Alfv\'en speeds)
in a fixed-radius cylindrical shell. Individual speeds are estimated as
$V_{\rm FF} \sim 0.97 \ c$, $V_{\rm FS} \sim 0.94 \ c$, $V_{\rm RS} \sim
0.73 \ c$, and $V_{\rm RF} \sim 0.67 \ c$, respectively. For a viewing
angle of $\theta \sim 14^{\circ}$ at {\em HST}-1 \citep[]{WZ09}, the
faster component {\em HST}-1c1 has $\sim 0.97 \ c$, while the slower
component {\em HST}-1c2 has $\sim 0.67 \ c$. As is mentioned in \S
\ref{sec:Model}, a separation of observed super/sub-luminal components
can be identified as distinct proper motions of two fast-mode MHD
shocks, $V_{\rm FF}$ and $V_{\rm RF}$. Therefore, our numerical model is
reasonably consistent with observations \citep[]{C07} at a
quantitative level.

We performed a simulation with an underlying velocity (the left state in
the Riemann problem we solve) that was constrained by the maximum
velocity of the {\em HST}-1d ($ < 0.25c$) allowed by the observations.
There is no evidence for an additional underlying flow other than the
velocity of the {\em HST}-1d component itself. That is, the flow that
produces the superluminal motion of {\em HST}-1c1 is the flow, and that
is our model: the ejection of a new MHD jet from HST-1 at a speed much
greater than any underlying flow is what produces the recent
observations of HST-1 complex suggested by \citet[]{C07}. The FF shock
that mainly determines the Lorentz factor for the HST-1 complex is {\em
HST}-1c1. What we learn from the observations is that the RF shock ({\em
HST}-1c2) is subluminal; a property predicted by the quad shock
model. This constitutes a constraint despite the fact that the exact
speed of the RF shock is determined by the model parameters which have
been specified only within a given range. However, we emphasize that a
very wide range of those parameters predicts sub-luminal motion for the
RF shock.

In a forthcoming paper, we will study the propagation of quad
relativistic MHD shocks in a conical geometry ($z \propto r$) with an
increasing cross section that is compatible with the M87 jet on
$10^{2-3}$ pc scales. This will allow us to study deceleration of the
shocks as the jet propagates on $10^{2-3}$ pc scales \citep[]{B95, B99}.

\section{Discussion and Conclusions}
\label{sec:Discussion} 

\begin{figure} 
\begin{center}
\includegraphics[scale=0.18, angle=0]{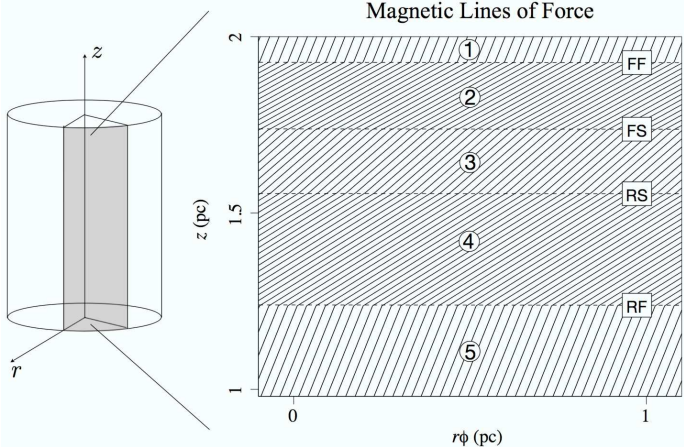} \caption{ \label{fig:f4}
Variation of the helical magnetic field in the $\phi$-$z$ plane
($B_{\phi},\ B_{z}$) and in the frame of the galaxy, as the jet flow (in
$z$) crosses the quad MHD shocks (labeled at right). (Compare the
right-hand panel with Fig. \ref{fig:f3}e.)  Only $\sim 1$ radian of the
pattern is shown, which, of course, is periodic over the full $2 \pi$
radian circumference of the cylinder. The pitch angles of the field
upstream of the RF and downstream of the FF (both are
``shock-upstream'') are equal (in this frame). Changes in the pitch
angle are caused by jump conditions at the different shocks. Each state
between quad MHD shocks is identified by numbers}
\end{center}
\end{figure}

Figure \ref{fig:f4} shows the variations in the magnetic helical lines
of force over $\sim 1$ radian of the $\phi$-$z$ plane as the jet
(propagating in the $z$ direction) passes through each of quad MHD
shocks. Magnetic field strengths $|\bolB|$ in the regions upstream of RF
($\textcircled{\footnotesize 5}$) and downstream of FF
($\textcircled{\footnotesize 1}$) are about 10.9 mG ($B_{\phi}/B_{z}
\simeq 0.33$). As mentioned above, the azimuthal field components in the
inter-shocked regions are amplified by compression, and they dominate
over the axial component {\em in the rest frame of the galaxy};
$B_{\phi}/B_{z}$ and $|\bolB|$ are $\sim 2.3$ and $\sim 25.7$ mG (FS --
FF: $\textcircled{\footnotesize 2}$), $\sim 1.2$ and $\sim 16.0$ mG (RF
-- RS: $\textcircled{\footnotesize 4}$), respectively.  On the other
hand, {\em in the rest frame of the fluid element}, only the forward
part of inter-shocked region (FS--FF: $\textcircled{\footnotesize 3}$)
has large magnetic pitch $(B_{\phi}/\gamma)/B_{z} \sim 1.3$.

One of the important insights that we derive from our numerical model is
an asymmetric structure of inter-shocked regions such as gas compression
and magnetic pitch angle $\theta_{\rm pitch}^{'} [\equiv
\tan^{-1}(B_{\phi}/\gamma/B_{z})]$ (in the rest frame of the fluid
element) in pre-shock regions of the two fast shocks: $\theta_{\rm
pitch}^{'} \sim 17^{\circ}$ {\em down}stream of the FF
($\textcircled{\footnotesize 1}$), and $\sim 3^{\circ}$ {\em up}stream
of the RF ($\textcircled{\footnotesize 5}$) ($\theta_{\rm pitch}$ in the
frame of the galaxy is same at the downstream of the FF and the upstream
of the RF, as is shown in Fig. 4). Efficiencies and mechanisms of
relativistic particle acceleration depends crucially on both the
magnetization and the magnetic obliquity of the upstream
plasma. Particle acceleration is mostly mediated by DSA for
quasi-parallel shocks ($\theta_{\rm pitch}^{'} \lesssim 10^{\circ}$),
but shock drift acceleration (SDA) is the main acceleration mechanism
for larger magnetic obliquity \citep[]{SS09}. In the rest frame of the
fluid element, the density increases by $r_{\rm cmp.} \simeq 3.42$ at
the RF and $r_{\rm cmp.} \simeq 1.90$ at the FF (where $r_{\rm cmp.}
\equiv \rho_2/\rho_1$ is the shock compression ratio and $\rho_{1,\, 2}$
are the density ahead and behind the shock). The slope $\delta$ in the
particle energy distribution for DSA is determined as $\delta=(r_{\rm
cmp.}+2)/(r_{\rm cmp.}-1)$ \citep[]{B78} and thus $\delta \simeq 2.23$
is expected at the RF, which agrees to the theory and observation
described in \S \ref{sec:Model}. In a forthcoming paper, we will examine
these aspects (DSA/SDA) in details to identify observed properties of
the M87 jet in $10^{2-3}$ pc scales as a context of quad relativistic
MHD shock system.

As mentioned in \S \ref{sec:Model}, the flow downstream of {\em HST}-1
has a conical structure with a constant opening angle until it reaches
knot A. Poloidal magnetic field varies as $B_{z} \propto r^{-2}$ in this
region, while toroidal magnetic field varies as $B_{\phi} \propto
r^{-1}$. With measured radii $r_{0} \sim 1$ pc near {\em HST}-1d
\citep[]{C07} and $r_{1} \sim 33.2 - 44.1$ pc at knot A \citep[]{O89,
S96}, combined with the $\bolB$-field strengths ($B_{\phi} \sim 12.39$
mG and $B_{z}\sim 10.32$ mG) at the reverse feature (RF -- RS:
$\textcircled{\footnotesize 4}$) in the simulation, we estimate $B_{\phi
1} \sim 280.9-373.2$ $\mu$G and $B_{z 1} \sim 5.3-9.4$ $\mu$G as a
counterpart of knot A. This strongly toroidally dominated field agrees
with the polarization observations \citep[]{O89, P99}.  Using the upper
limits on inverse Compton radiation imposed by the HESS and HEGRA
Cerenkov telescope observations of the kpc scale jet to estimate the
magnetic field strength in the brightest knot A, \citet[]{S05} obtain a
``safe'' lower limit of $|\bolB| \gtrsim 300$ $\mu$G, indicating a
departure from the equipartition value (from the synchrotron spectrum of
the knot A); the magnetic field energy density within the brightest knot
is very likely higher than the energy density of the radiating
ultrarelativistic electrons. This implies that interknot regions are
likely to be extremely magnetized. On the other hand, a reasonable upper
limit to the field strength of knot A can be constrained by the total
power $L_{\rm j}$ of the M87 jet from the relation $\pi r_{1}^2 (\bolV
\times \bolB) \times \bolB / (4 \pi) \leq L_{\rm j}/2$, where $L_{\rm j}
\sim$ few $\times 10^{44}$ erg s$^{-1}$ \citep[]{O00}, and we assume an
equipartition between the matter energy and Poynting fluxes. In a
situation in which $V_{z} \simeq c > V_{\phi} \gg V_{r}$, $B_{\phi} \gg
B_{z} > B_{r}$, the dominant term in the axial ($z$) direction, the main
carrier of electromagnetic energy, is $B_{\phi}^2 V_{z}/(4 \pi) \simeq
B_{\phi}^2 c/(4 \pi)$. This gives $|\bolB| (B_{\phi}) \leq 1$ mG. Thus,
these constraints are consistent with our derivation $|\bolB| \sim
289-370$ $\mu$G at around knot A by considering an expansion of jet
cross section.

Sideways oscillations of the brightness ridge-lines downstream of the
{\em HST}-1 complex are closely followed by the filamentary structure of
distributed magnetic fields \citep[]{O89}. Some of these filaments can
be interpreted as a 3-dimensional helix seen in projection, produced by
growing current-driven instabilities (CDIs) \citep[]{N01}.  Indeed,
\citet[]{NM04} found that MHD shocks play an important role in
triggering helical kink CDIs: while rotation of the plasma may stabilize
the helical kink instability locally, MHD shocks (particularly in the
region A -- B -- C) can rapidly alter this stabilizing rotation (Figure
\ref{fig:f3}), suddenly causing the magnetized plasma to violate the
Kruskal-Shafranov stability criterion. Further downstream of knot C, the
knot-like features disappear, and the jet becomes diffusive with strong
side-to-side oscillations and bending. This may also be a good example
of the growing ``external mode of the CDIs beyond the X-ray cluster core
where a separation of the paths between the jet forward and
return current occurs due to the rapidly decreasing density of the
external thermal gas'' \citep[]{N07}. We shall examine the above topics
in forthcoming papers by treating our quad MHD shock model in 1-D
conical and, eventually fully 3-D, flow environments.

The idea that magnetic fields are instrumental in the formation and
propagation of jets in active galactic nuclei dates back four decades.
Despite a recent growing consensus on this notion stemming from the
results of numerical simulations of magnetohydrodynamic flows near
black holes, the precise dynamical role of magnetic fields in observed
parsec and kiloparsec jets has remained uncertain. In this first in a series
of papers, we have combined SRMHD numerical simulation with observation
to explain superluminal knot ejections from {\em HST-1} in the M87
jet. Indeed this is the first examination of superluminal motions in the
extragalactic jet as a consequence of trailing MHD shocks in a
relativistic flow that possesses helically twisted magnetic
structures. We begin to develop a detailed picture for the jet in M87
that is grounded in the dynamics of relativistic MHD, ultimately
suggesting that the entire jet (from subparsec to kiloparsec scales) is
an MHD phenomenon. Eventually our model may also be readily applicable
to other jet systems, and it may constitute the foundation of an MHD
paradigm for astrophysical jets in general.

Stimulating discussions with Colin A. Norman, Keiichi Asada, and Jose
Gracia are gratefully acknowledged. M.N. is supported by the Allan C.
Davis fellowship jointly awarded by the Department of Physics and
Astronomy at Johns Hopkins University and the Space Telescope Science
Institute. Part of this research described was carried out at the Jet
Propulsion Laboratory, California Institute of Technology, under
contract to the National Aeronautics and Space Administration. D.G. is
supported by the NASA Postdoctoral Program at NASA JPL administered by
Oak Ridge Associated Universities through contract with NASA.


\end{document}